\documentclass[9pt,twocolumn,twoside]{pnas-new}
\templatetype{pnasresearcharticle}

\title{Solidification and superlubricity with molecular alkane films}

\author[a,b,1]{Alexander M. Smith}
\author[a]{James E. Hallett} 
\author[a,1]{Susan Perkin} 

\affil[a]{Department of Chemistry, Physical and Theoretical Chemistry Laboratory, University of Oxford, Oxford OX1 3QZ, U.K.}
\affil[b]{Department of Inorganic and Analytical Chemistry, University of Geneva, 1205 Geneva, Switzerland}

\leadauthor{Perkin} 

\significancestatement{An essential step in society's attempt to reduce energy consumption is to understand and optimize energy dissipation (friction) in mechanical processes. Despite decades of study, one of the most fundamental model systems in the study of friction remains unsolved: the shear of molecular films of simple alkanes between atomically smooth crystalline surfaces. Previous reports of the friction coefficient diverge over three orders of magnitude and reports that the film solidifies on confinement remain controversial. We perform measurements with control over parameters not previously taken into account, allowing us to reconcile earlier reports. We demonstrate that the alkane films indeed solidify, and yet retain ultra-low friction coefficients under certain conditions.}

\authorcontributions{AMS carried out the experiments. AMS and JH carried out analysis of the data.  AMS and SP together designed the project and interpreted the experiments. All authors contributed to drafting the manuscript.}
\authordeclaration{The authors declare no conflicts of interest.}
\correspondingauthor{\textsuperscript{1}To whom correspondence may be addressed. E-mail: susan.perkin@chem.ox.ac.uk, Alexander.Smith@unige.ch}


\begin{abstract}
Hydrocarbon films confined between smooth mica surfaces have long provided an experimental playground for model studies of structure and dynamics of confined liquids. However fundamental questions regarding the phase behavior and shear properties in this simple system remain unsolved. With ultra-sensitive resolution in film thickness and shear stress, and control over the crystallographic alignment of the confining surfaces, we here investigate the shear forces transmitted across nanoscale films of dodecane down to a single molecular layer. We resolve the conditions under which liquid-solid phase transitions occur and explain friction coefficients spanning several orders of magnitude. We find that commensurate surface alignment and presence of water at the interfaces each lead to moderate or high friction, whereas friction coefficients down to $\mu \sim0.001$ are observed for a single molecular layer of dodecane trapped between crystallographically misaligned dry surfaces. This ultralow friction is attributed to sliding at the incommensurate interface between one of the mica surfaces and the laterally ordered solid molecular film, reconciling previous interpretations. 
\end{abstract}

\dates{This manuscript was compiled on \today}
\doi{\url{www.pnas.org/cgi/doi/10.1073/pnas.XXXXXXXXXX}}

\begin{document}

\maketitle
\thispagestyle{firststyle}
\ifthenelse{\boolean{shortarticle}}{\ifthenelse{\boolean{singlecolumn}}{\abscontentformatted}{\abscontent}}{}

\dropcap{T}he dynamics of nano-confined liquids are central to processes such as filtration, energy storage, lubrication, cell membrane pumps and channels, and a multitude of other technological and biological scenarios. When the confinement lengthscale is comparable to the molecular size of the contained liquid a spectrum of dynamic phenomena are possible depending on the detailed intermolecular and surface interactions at play\cite{Klein1995,Raviv2001,Smith2013,Bonthius2014,Secci2016,Comtet2017a}. In light of such strong relevance it is surprising that one of the simplest model systems of confined fluid -- a non-polar molecular liquid confined between two atomically smooth surfaces -- still evades full explanation. It is frequently reported that simple molecular liquids confined to films below 10~nm or so undergo a transition to solid-like behaviour\cite{VanAlsten1988,Gee1990,Klein1995,Klein1998,Demirel1996,Kumacheva1998} however the nature of this transition continues to motivate discussion\cite{Zhu2004,Gourdon2006,Kienle2016,Xu2018}. Furthermore, reported measurements of the shear stress sustained by these films span a wide spectrum with friction coefficients ranging from exceptionally low (for squalane\cite{Zhu2004}), through intermediate values ($\mathrm{\mu \sim} 0.1$) for linear alkanes\cite{Yoshizawa1993,Lundgren2008}, and even much higher values ($\mathrm{\mu > 1}$)\cite{Gee1990}; all measured using similar apparatus employing a single-asperity contact. The tantalising reports of \textit{superlubricity} in some cases, arbritrarily defined by a friction coefficient $\mathrm{\mu \lesssim 0.01}$, provides strong motivation to interrogate this model system to untangle and explain the array of observed phenomena \cite{Zhu2004,Ma2015,Cihan2016,Bhushan1995,Vanossi2013}.
 
Here we report measurements of the normal and lateral forces sustained by nanometric films of dodecane confined between two crystalline mica surfaces. We show that by tuning two parameters not controlled in previous works -- the degree of commensurability between the confining crystal planes and the ambient humidity -- we are able to span the range of previously-observed behaviors from superlubric to high friction. Ultralow friction across a single molecular layer of dodecane is possible when the two mica crystal surfaces are incommensurate and sufficiently dry.  Twisting the crystalline confining surfaces into commensurate orientation leads to higher friction, as does the presence of water adsorbed on the surfaces. These observations imply strategies for achieving ultra-low friction under ambient conditions, and indicate a resolution to the longstanding debate about friction across nonpolar molecular films. 

\begin{figure}%[tbhp]
\centering
\includegraphics[width=.8 \linewidth]{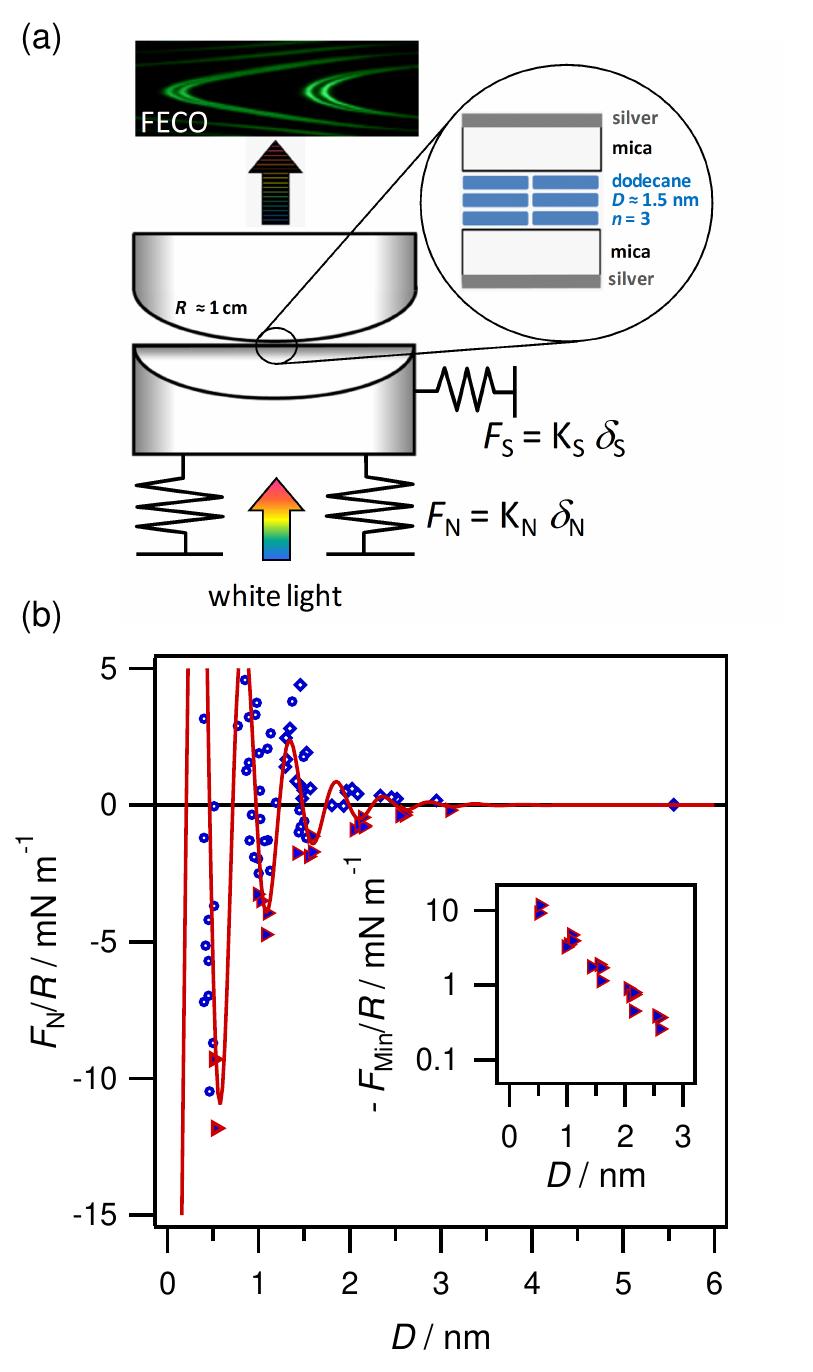}
\caption{(a) Schematic diagram of the SFB setup, showing the path of light through the optical lenses arranged in crossed-cylinder orientation. A photograph of a typical interference spectrum (FECO), used to determine $D$ and precise contact geometry, is shown above. A zoom-in to the contact area shows detail of the interferometric cavity with liquid confined to nanofilms between two atomically smooth mica sheets. (b)Measured normal force, normalized by the radius of curvature, between two mica sheets arranged in crossed-cylinder configuration across a dodecane film as a function of dodecane film thickness. Diamonds show data measured on approach of the surfaces, circles show data measured on retraction, and adhesive minima are marked as filled triangles. The solid line fits the measured force to $F_N/R = Ae^{-D/\lambda_o}cos(\omega D+\phi)$, with $2 \pi /\omega =$ 0.49 nm, $\lambda_o =$ 0.51 nm (further discussion in the Supplementary Information). The upper inset shows the crossed cylinder configuration of the mica surfaces; $D$ is measured at their point of closest approach. The lower inset shows the magnitude of the adhesive minima on a log-linear scale. Data in this figure correspond to mica sheets in crystallographic alignement ($\theta = 0^\circ$); a comparison with $\theta = 90^\circ$ is provided in the SI Appendix.}
\label{Fig:F1}
\end{figure}

Our measurements are carried out using a Surface Force Balance (SFB) to measure normal and shear forces transmitted between the mica surfaces across thin films of dodecane. The two mica sheets are freshly-cleaved along the crystal plane to yield atomically-smooth and uniform sheets (no steps in the crystal top or bottom) over $\sim$1 cm$^2$ area and with thickness 1-4 $\mu$m. The mica crystals are back-silvered and mounted upon hemi-cylindrical lenses (radius of curvature $R$ = 7-10 mm), which are then arranged inside the SFB in crossed-cylinder orientation. Dodecane is injected as a droplet (~50 $\mu$l) suspended between the lenses. Collimated white light is directed normal to the mica-liquid-mica optical cavity between the two silver layers, and the resulting interference pattern (Fringes of Equal Chromatic Order, FECO) are used to determine the dodecane film thickness to within 0.1 nm, after the mica thickness has been determined by calibration. The lenses are translated normally and laterally relative to one another using a piezo crystal upon which one lenses is mounted. Normal and lateral forces are determined from the deflection of springs, whose spring constants are calibrated separately. The apparatus and procedures have been described in more detail elsewhere\cite{Klein1998,Perkin2006} and some further details particular to the present measurements are provided in the SI Appendix. A schematic of the geometry and key features is shown in Figure \ref{Fig:F1}(a). 

Mica was highest grade of the ruby muscovite variety (S\&J Trading Inc.) . In some experiments, monolayers of octadecyltriethoxysilane (OTE) (Fluorochem, 95\%) were deposited on mica according to established procedures\cite{Malham2009}; these surfaces are amorphous with roughness similar to mica. Dodecane (Sigma-Aldrich, anhydrous, $>$99\%) was stored over freshly activated molecular sieves for several days, and filtered through a 0.2 $\mu$m PTFE membrane immediately prior to injection of the droplet.  In most experiments reported here, mica was re-cleaved with tape using the method of Frantz and Salmeron\cite{Frantz1998}, and immediately mounted in the apparatus to minimise exposure to laboratory air. With this method the mica surfaces are typically exposed to particle-free lab air for $\sim$ 30 seconds before sealing into the dry and inert atmosphere inside the SFB chamber. The chamber is purged with dry nitrogen and a dish containing P$_2$O$_5$ is present to absorb any residual water vapour. When this re-cleaving method is used the mica-film-mica 3-layer interferometric cavity becomes weakly asymmetric. To calculate $D$ in this setup a multi-matrix fitting procedure can be used, although with films $D \leq 10$nm the results are indistinguishable from those obtained with the analytic expression for a symmetric interferometer and so this was used in our experiments. An extended discussion of these methods and a comparison of data fitting with the matrix calculation and symmetric analytic equation are provided in the SI.
	
	Central to the technique is the ability to resolve the thickness of the liquid between the surfaces, $D$, with sub-molecular resolution using white-light multiple beam interferometry\cite{Israelachvili1973} with the aid of semi-reflective silver mirrors on the backside of each mica piece. Hence normal and shear forces can be measured between the surfaces \textit{via} the deflection of horizontal and vertical springs -- of known spring constants -- with simultaneous knowledge of the liquid film thickness and contact geometry.  Furthermore, exploiting the birefringence of mica, the multiple-beam interference spectrum can also be used to determine the relative twist angle of the two mica sheets, \textit{in situ}, from the doublet splitting of each interference fringe. Crystallographically aligned mica ($\mathrm{\theta} = 0^\circ$) results in doublets with the maximum possible splitting, while mica completely misaligned ($\mathrm{\theta} = 90^\circ$) results in zero splitting and the fringes appearing as perfect singlets in the interference pattern. In this work, experiments were carried out at varying twist angles using mica cleaved from the same original crystal facet then aligned at the desired angle by gluing to the cylindrical lenses at appropriate orientations. The error associated with alignment in this way depends on the observation of doublets: for crystallographically aligned surfaces ($\mathrm{\theta} = 0^\circ$) the twist angle can be determined very accurately from the doublet splitting of the FECO fringes in the interference spectrum, with an error $\pm$1$^\circ$. However when $\mathrm{\theta}$ is close to $90^\circ$, the exact twist angle cannot be determined due to the fringes being superimposed (appearing as a singlet). We overcame this problem by carrying out series of experiments using mica cleaved from the same original crystal facet then aligned at the desired twist angle by gluing to the cylindrical lenses at appropriate orientations. In this way, experiments at $\mathrm{\theta} = 90^\circ$ can be achieved by mounting the mica crystals onto lenses after the lattice orientation was known precisely from another experiment with mica from the same facet. The pieces are cut with straight edges before gluing to facilitate exact misaglinment when viewed through the microscope objective. This method leads to a larger total aligment error of ±5$^\circ$ for misaligned surfaces. Therefore, in the following, `$\mathrm{\theta} = 0^\circ$' implies $0^\circ \pm 1^\circ$, and `$\mathrm{\theta} = 90^\circ$' implies $90^\circ \pm 5^\circ$. 
    
\begin{figure*}%[tbhp]
\centering
\includegraphics[width=.65\linewidth]{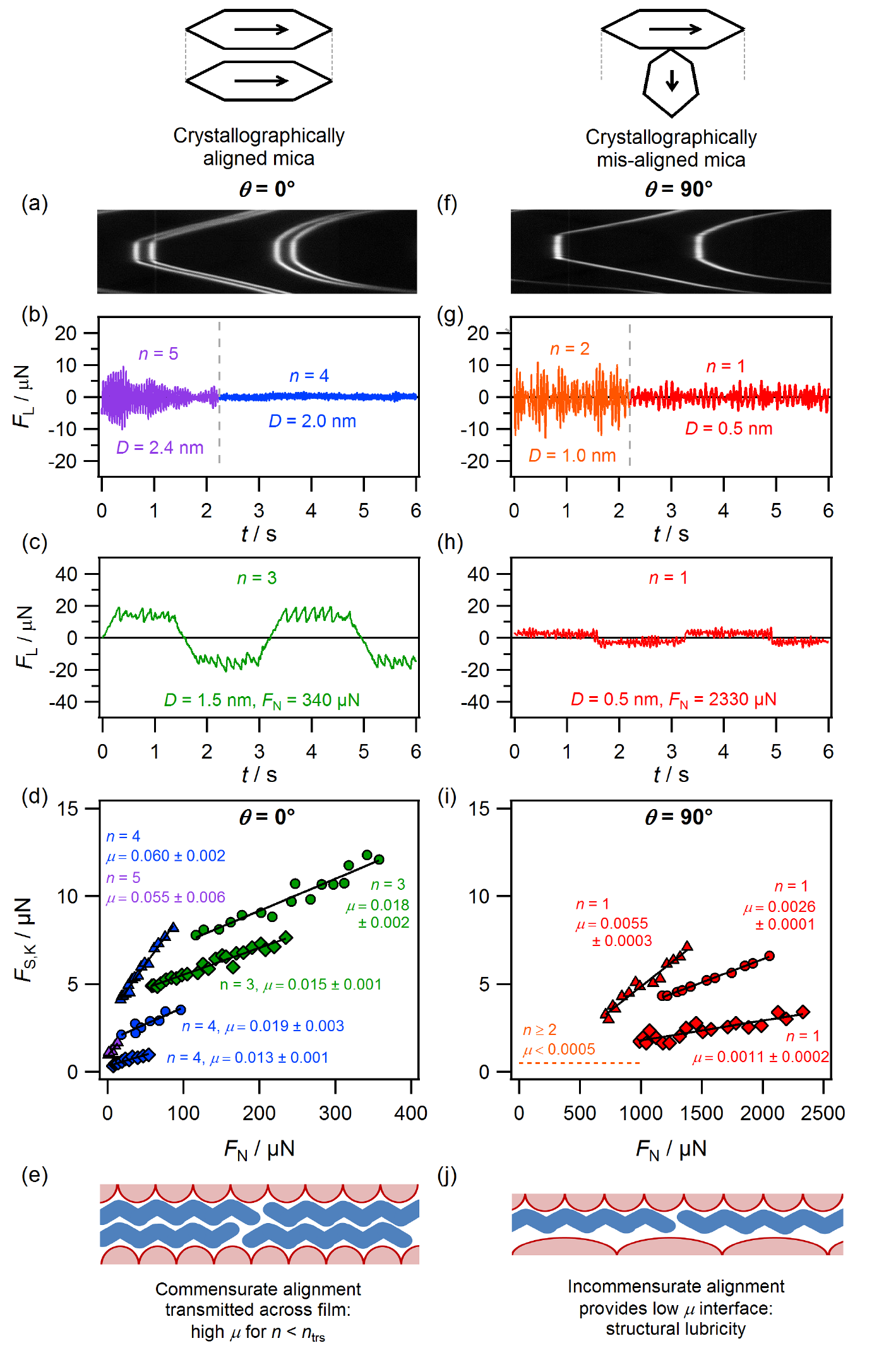}
\caption{Comparison of lateral force measurements carried out with dodecane films between two mica surfaces mounted in relative crystallographic alignment, $\theta = 0^\circ$ (a)-(e), and in crystallographic misalignment, $\theta = 90^\circ$ (f)-(j). The extent of crystallographic alignment is determined from the FECO interference pattern as described in the text; panels (a) and (f) show example FECO patterns for aligned and misaligned surfaces. Panels (b) and (g) show examples of the lateral fluctuations in force arising from interaction with the ambient thermal and mechanical environment, captured during the transition from $n=(n_{trs}+1)$ to $n=n_{trs}$, i.e. the point at which the film sharply steps from being liquid-like to solid-like. For the example of aligned surfaces in (b) this occurs from 5 to 4 dodecane layers, whereas for misaligned surfaces, in (g), this always occurs between $n = 2$ to $n = 1$. Panels (c) and (h) show typical traces of the lateral force transmitted across the film when one surface is subjected to back-and-forth sliding motion at constant velocity $190 \mathrm{nm s^{-1}}$ and amplitude 300 nm. For all films below the transition to solid-like, i.e. $n \leq n_{trs}$, a finite stress is sustained by the film before sliding occurs. For aligned surfaces (in (c)) the shear stress is high, and after the yield point we observe a series of stick-slip cycles until the direction of shearing is reversed. For misaligned surfaces (in (h)) the lateral force sustained by the $n = 1$ film is very small, and after yielding we observe smooth sliding (within our resolution). In panels (d) and (i) we show how the shear stress sustained by the films - \textit{i.e.} the kinetic friction force $F_{S,K}$ - varies with $F_{N}$ for a range a different $n$ all below $n_{trs}$ and from various experiments to show the variation between experiments with different mica sheets.  The gradient of $F_{S,K}$ with $F_{N}$ is the friction coefficient, $\mu$, and is shown inset to the panels for each series. Different symbols correspond to six independent experiments each using different mica pieces and so demonstrate the experimental variation.  In (e) and (j) we provide schematics to illustrate our interpretation, as described in the text: a laterally-ordered dodecane film can sit in commensurate alignment with the mica surface, transmitting the mica crystalline symmetry into the film. In the case where the two mica sheets are crystallographically aligned, (e), the solidified and 2D-ordered dodecane film can cause the surfaces to 'lock' together and the junction exhibits higher friction. Conversely, when the mica sheets are misaligned, (j), such that their lattices are incommensurate, the dodecane film can only align with one (not both) surfaces and there must always exist at least one interface of lattice mis-match and therefore low friction.}
\label{Fig:F2}
\end{figure*}

We begin by describing the measured normal force between the mica sheets across dodecane as a function of film thickness, Figure \ref{Fig:F1}. As the surfaces approach from large separations they first experience a weak (Van der Waals) attraction, measurable from $\sim$5 nm, then at a distance of 3 nm there is an abrupt repulsion and the force profile becomes oscillatory at $D <$ 3 nm with a wavelength of 0.5 nm. As the surfaces approach they experience repulsive regions of the force law; retraction of the surfaces from any point in the oscillatory region leads to exploration of the minima and an eventual jump-apart of the surfaces from the nearest adhesive minimum (marked as filled triangles in Figure \ref{Fig:F1}) to larger separations. The magnitudes of adhesive minima increase exponentially with decreasing film thickness down to one confined molecular layer as shown in the inset to Figure \ref{Fig:F1}. Such oscillatory force profiles, called structural forces, are commonly observed for molecular liquids which form ordered layers between two surfaces\cite{Horn1980,Horn1981,Christenson1983,Christenson1987}. The measured oscillatory wavelength, $\sim$0.5 nm, corresponds to the width of the molecule rather than its length or any other average dimension, indicating a parallel orientation of the linear alkane molecules near the surfaces in agreement with previous force measurements with even-numbered alkanes from hexane to hexadecane\cite{Christenson1987,Gosvami2007}. In the present measurements six oscillations in the force profile are apparent, corresponding to up to six layers of dodecane molecules confined between the mica surfaces. The strong repulsive barrier at $\sim$0.5 nm and resultant flattening of the surfaces prevented squeeze-out of the last molecular layer.  Notably, we find here that the wavelength and magnitude of the structural forces are insensitive to the relative orientation of the mica sheets, or even the confining material itself (see Supplementary Information, Figure S1).  
	 
We now turn to measurements of the shear force sustained by the film with varying number of dodecane layers, \textit{n}. Some representative examples are shown in Figure \ref{Fig:F2}, where panels (a)-(e) on the left correspond to crystallographically aligned mica ($\mathrm{\theta} = 0^\circ$) and panels (f)-(j) on the right correspond to experiments with misaligned mica at $\mathrm{\theta} = 90^\circ$. 
In all measurements we routinely observe a sharp transition from pure viscous (liquid-like) to elastic (solid-like) shear response of the dodecane film when the film thickness is decreased to a certain number of layers, which we label $n_{trs}$. The transition is marked by (i) a sharp reduction in the ambient noise (increase in damping) lateral to the surfaces (Figure \ref{Fig:F2}b, g) and (ii) onset of finite yield stress sustained by the film (Figure \ref{Fig:F2}c, h). This is in accordance with confinement induced liquid-to-solid phase transitions first observed for confined pseudo-spherical molecules such as cyclohexane and OMCTS\cite{Klein1995,Klein1998}. The value of $n_{trs}$ depends on a variety of factors including twist angle, ambient vibrations, and the presence of water at the surfaces, as follows.   

	For crystallographically aligned mica surfaces ($\mathrm{\theta} = 0^\circ$; Figure \ref{Fig:F2} a-e), shearing the surfaces laterally across this solid-like confined film, i.e. for films of $n$ layers with $n \leq n_{trs}$ (with $4 < n_{trs} < 8$  when $\mathrm{\theta} = 0^\circ$ in our experiments), results in stick-slip friction as shown in Figure 2c and as has been observed for molecularly confined films of hexadecane in the past\cite{Yoshizawa1993,Yoshizawa1993b}. During stick-slip sliding motion, the lateral force varies between a maximum static shear force $F_{S,st}$, at the point of slip, and a minimum kinetic shear force, $F_{S,k}$, at the stick-point at which the surfaces are coupled together again. This behavior continues until the sliding direction is reversed.  The variation of $F_{S,k}$ with load for films of n = 3, 4 and 5 is shown in Figure \ref{Fig:F2}(d) for different experiments each using mica from different crystal facets. We choose to plot the kinetic friction force, $F_{S,k}$, as it remains constant over the range of shear velocities investigated, whereas $F_{S,st}$ was observed to increase with decreasing velocity as as been well documented in the past for similar systems \cite{Drummond2001,Urbakh2004} 
	
	The situation is dramatially different when the mica surfaces are arranged in crystallographic misalignment ($\mathrm{\theta} = 90^\circ$; Figure \ref{Fig:F2} f-j). With misaligned mica sheets $ n_{trs} =1$: the transition to solid-like behaviour occurs at a distance of 0.5 nm, corresponding to only one molecular layer, at which point there is a small but measurable yield point and kinetic friction force. The surfaces slide smoothly across the layer with no observable stick-slip cycles above our force resolution. Friction coefficients for the misaligned surfaces are extremely low (Figure \ref{Fig:F2}i), far into the 'superlubric' regime, and generally an order of magnitude lower than for aligned surfaces as seen by comparing Figure \ref{Fig:F2} d and i. Within a single experiment (same mica sheets), friction behaviour was reproducible over different surface contact positions. Small differences between independent experiments with different mica pieces as shown in Figure \ref{Fig:F2}i may be due to differing amounts of adsorbed water, the small error in alignment of the surfaces, variation in sliding direction or likely a combination of these. Nevertheless the friction between misaligned surfaces is systematically observed to be an order of magnitude lower than between aligned surfaces. Importantly, the ultra-low friction reported here for misaligned surfaces can be achieved using mica cleaved and cut with Pt wire using standard procedures with careful consideration of laminar flow\cite{Perkin2006}, and without the need for prior re-cleaving of the mica surfaces. However a low humidity in the lab is required ($<$20$\%$) for the duration of surface preparation (also see discussion in the SI Appendix). Without such precautions or prior re-cleaving, friction coefficients for dodecane confined between mica surfaces are typically in the range 0.1 $< \mu <$ 2.

\begin{figure}%[tbhp]
\centering
\includegraphics[width=.8\linewidth]{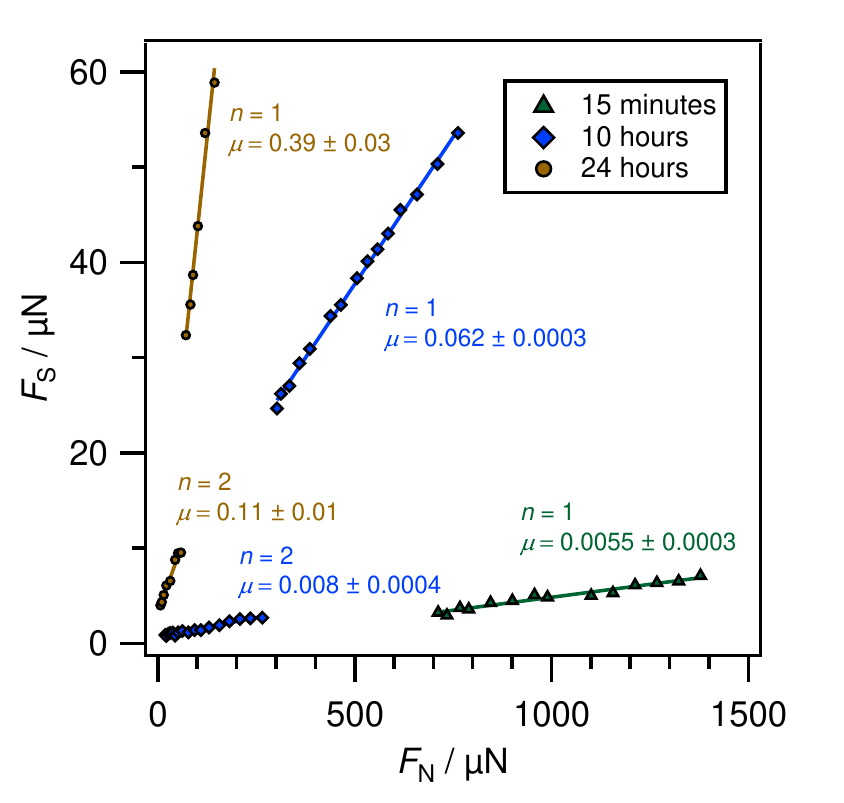}
\caption{Kinetic shear force as a function of applied normal load between crystallographically misaligned mica surfaces across dodecane films of one or two molecular layers, labelled $n = 1$ or $n = 2$ respectively, as a function of time exposed to air. The friction coefficient, $\mu$, across each layer is given by the gradient of each set of points.}
\label{Fig:F3}
\end{figure}

	 The presence of trace amounts of water is found to have a substantial effect on the friction behaviour, as shown in Figure \ref{Fig:F3} where we report friction as a function of time elapsed since exposure of the system to humid particle-free air. Here, the surfaces are crystallographically misaligned and friction is first measured as soon as possible after immersion in dodecane (15 minutes). Subsequently the dry nitrogen atmosphere inside the measurement chamber is exchanged with particle-free humid air. Friction is then recorded, \textit{with the same surfaces and at the same contact position}, after various air-exposure times. The friction increases very substantially: the friction coefficient for a single layer increases by almost two orders of magnitude over 24 hours, and the friction for two confined molecular layers becomes measurable, as water is gradually absorbed into the dodecane and diffuses to the interface.	
     
We now turn to discussing the likely mechanisms underlying the observed friction across nanofilms of dodecane, which we observe to span three orders of magnitude in friction coefficient depending on the crystallographic alignment of the confining surfaces and the water incorporation into the film. As a first step, we consider the likely phase and order in the confined films. As mentioned above, the clear discontinuity in mechanical properties of the dodecane film between at the point $n =  n_{trs}$ implies a change of state from liquid (purely viscous behaviour) at $n >  n_{trs}$ to a solid-like film (supporting finite shear stress) at $n \leq  n_{trs}$. This observation alone would be consistent with \textit{either} an amorphous solid / glass structure, \textit{or} crystalline order at $n \leq  n_{trs}$. However, from our measurements carried out at varying twist angle, we find enormously varying values of $n_{trs}$ and friction coefficient, $\mu$, as the confining surfaces go in and out of commensurate alignment. If the film were amorphous then the twist angle would not be expected to have any impact on  $n_{trs}$ or $\mu$, since the surface symmetry information would be `lost' in the amorphous intervening layer. Thus it appears that the symmetry of the crystalline surface is transmitted into the film, and across the film at $n \leq  n_{trs}$, implying some extent of crystalline or 2D ordered structure. 

The mechanism of stick-slip friction in molecular liquids has in the past been attributed to shear-melting of the solidified film\cite{Thompson1990}, however more recent experiments found no evidence for melting and both experiments and simulations point instead to an interfacial slip mechanism\cite{Robbins1995,RosenhekGoldian2015}. High resolution measurements using the non-polar liquid octamethylcyclotetrasiloxane (OMCTS) found no evidence of film dilation associated with the fleeting slip events during shear, which would be expected upon liquefaction of the thin film\cite{RosenhekGoldian2015}. This implies that interlayer slips within the film, or at the interface with the confining surfaces are more likely to be the dominant dissipation modes. Indeed, molecular dynamics simulations also reveal interlayer and wall slips during stick-slip sliding, where the ordered solidified film is well maintained during the slip\cite{Lei2011}.
	 	
		Comparing the friction measured for one and two molecular layers of dodecane (Figure \ref{Fig:F3}) allows us to distinguish between wall slip and interlayer slip mechanisms. For one confined dodecane layer, slip must take place at the mica interface. It is clear that friction coefficients are substantially different (lower) for a film of two confined layers, which suggests that in this case shear occurs between the dodecane layers - rather than at the wall - when the mica surfaces are misaligned. In the case of crystallographically aligned mica surfaces it is not clear at which interface slip is occurring since we are unable to compare with friction for one layer in that case. However we note that $\mu$ is essentially constant for all $n$ within each individual experiment (Figure \ref{Fig:F2}d), only the adhesion contribution (\textit{i.e.} the y-intercept of an extrapolated line through the points at each $n$; see SI Appendix for further discussion) alters with $n$, indicating that the slip plane is unchanged.
	
	The extremely low friction reported here for a single layer of ordered (2D crystalline) dodecane confined between dry crystallographically misaligned mica surfaces can be rationalised by \textit{structural lubricity} arguments  as follows. The concept of structural (super-)lubricity is well established for interpreting the low friction between dry crystalline surfaces at incommensurate alignment\cite{Hod2018}; the lattice mis-match between two crystalline surface leads to an effective `cancellation' of the energy barriers to lateral motion and thus ultra-low friction\cite{Hod2018}. However this concept was previously not thought relevant to liquid-lubricated contacts. Here we argue that, when the pure alkane film between crystalline surfaces solidifies into an ordered (2D crystalline) film, a similar structural lubricity can occur. The dodecane layers directly adsorbed to the mica surfaces are laterally ordered in domains such that they are in registry with the hexagonal mica lattice\cite{Docherty2010}. Hence, the hexagonal surface structure induces a hexagonal structure in the confined liquid. This order extends into the liquid over some characteristic distance, and, for aligned surfaces, can persist across the whole thickness of the film\cite{Jabbarzadeh2006} leading to measurable shear stress for  $n \leq n_{trs}$. However for misaligned surfaces there must always be an interface which is incommensurate and so allows ultralow friction, whether it be an interface between (within) the laterally ordered alkane layers or at the interface with one of the confining mica surfaces. Schematic diagrams illustrating this concept are in Figure \ref{Fig:F2}(e),(j). Such an effect has indeed been shown to occur in simulations of dodecane confined between crystalline surfaces\cite{Jabbarzadeh2007}, and cyclohexane between mica surfaces\cite{Xu2018} where the slip plane was unambiguously at the mica surface as is consistent with the present results. 
	
	To investigate the effect of surface crystallinity on the friction - testing the hypothesis above - we carried out separate experiments where the mica surface was coated with an atomically smooth but amorphous silane monolayer. In those experiments (examples are shown in the SI Appendix) we found no effect of twist angle, highlighting the role of crystallinity of the confining surfaces. In these experiments with amorphous surfaces the kinetic friction forces were reproducibly low, though the static friction force increased markedly with load. It is likely that the amorphous surfaces prevent pinning of local domain alignment compared to crystalline surfaces\cite{Jabbarzadeh2006b}.
	
	While the effect of surface orientation on friction across hexadecane has been studied experimentally\cite{Charrault2013}, the friction reported was in all cases generally very high; this was perhaps due to small amounts of surface adsorbed water which we have here shown to increase friction. Whist we cannot unambiguously determine the mechanism of the increased friction with water content, there is strong evidence in the literature of an effect particular to mica which we propose as most likely to underlie our observations. It is well established that mica cleaved in a humid atmosphere adsorbs water and CO$_2$ which complex with the mica K$^+$ to form crystalline K$_2$CO$_3$ hydrates\cite{Christenson2016}. Since the water and carbonate crystal are insoluble in dodecane, it is likely that -- in the `wet' experiments where these crystals are allowed to form -- they remain on the mica surfaces. Since the crystals can easily reach dimensions of order 1-2nm, they can bridge the gap between the two mica sheets and thereby increase the shear stress of the contact region. The gradual increase in friction with increasing humidity is likely to correspond to a gradual increase in both the number and size of these crystalline salt bridges within the confined region. Notably, even in relatively humid experiments, the oscillatory normal force between the surface is not much different to the dry experiments; this is likely due to the very small fraction of the total area taken up by the crystallites, so that that they impact significantly the shear force but not the normal force. 
	
	In summary, we investigated the effects of surface crystallinity, crystallographic orientation, and presence of water on the friction across molecularly confined dodecane films. The measured oscillatory structural forces across alkanes reported here and in the literature are in agreement in both range and magnitude, and thus we propose that an explanation for the enormous differences in observed friction behaviour - spanning some three orders of magnitude - is likely to come from differences in lateral structure in the confined film. Based on the experiments presented here we propose that this lateral structure is acutely sensitive to the crystallinity and twist angle of the confining surfaces, as well as the presence of trace amounts of water, and that sliding of incommensurate solid-like alkane layers explains the ultralow ('superlubric') friction observed in certain circumstances. Our results reconcile the confinement-induced liquid-to-solid phase transition with ultralow friction observed in the past by different researchers, which until now have appeared to be competing findings.

\matmethods{Please describe your materials and methods here. This can be more than one paragraph, and may contain subsections and equations as required. Authors should include a statement in the methods section describing how readers will be able to access the data in the paper. 

\subsection*{Subsection for Method}
Example text for subsection.
}

%\showmatmethods{} % Display the Materials and Methods section

{\bf Data Availability}\\
All raw data and data protocols are available upon reasonable request to the authors. \\

\acknow{AMS is grateful for a Doctoral Prize from the EPSRC. SP is supported by The Leverhulme Trust (RPG-2015-328) and the European Research Council (under ERC Starting Grant no. 676861)}

\showacknow{} % Display the acknowledgments section

% Bibliography
\bibliography{dodecane_refs}

\end{document}